\documentclass[a4,12pt, amscd, amsfonts]{amsart}
\usepackage{amscd}
\usepackage{times}
\usepackage{amssymb,epic,eepic,epsfig,amsbsy,amsmath}
\usepackage[headings]{fullpage}
\numberwithin{equation}{section}
\newtheorem{thm}{Theorem}[section]
\newtheorem{defn}[thm]{Definition}
\newtheorem{cor}[thm]{Corollary}

\newtheorem{lem}[thm]{Lemma}
\newtheorem{rem}[thm]{Remark}

\def\RREF{\mathop{\mathrm {RREF}}\nolimits}
\def\FFT{\mathop{\mathrm {FFT}}\nolimits}
\def\add{\mathop{\mathrm {add}}\nolimits}
\def\inverse{\mathop{\mathrm {inverse}}\nolimits}
\def\H{\mathop{\mathrm {H}}\nolimits}
\def\U{\mathop{\mathrm {U}}\nolimits}
\def\rk{\mathop{\mathrm {rk}}\nolimits}
\begin{document}
\title[Quantum Gau\ss\ Jordan Elimination]{Quantum Gau\ss\ Jordan Elimination}
\author[Do Ngoc Diep and Do Hoang Giang]{Do Ngoc Diep${}^1$ and Do Hoang Giang${}^2$}
\date{}
\maketitle
\begin{abstract}
In this paper we construct the Quantum Gau\ss\ Jordan Elimination
(QGJE) Algorithm and estimate the complexity time of computation
of Reduced Row Echelon Form (RREF) of an $N\times N$ matrix using
QGJE procedure. The main theorem asserts that QGJE has computation
time of order $2^{N/2}$.
\end{abstract}
\section{Introduction}
Let us consider a general system of linear equations
\begin{equation}\left\{\begin{array}{ccc}
a_{11}x_1 + \dots +a_{1n}x_n & = & b_1\\
\dotfill &\dotfill & \dotfill\\
a_{m1}x_1 + \dots + a_{mn}x_n & = & b_m
\end{array}\right.\end{equation}
Denote
\begin{equation}A = \left[\begin{array}{ccc}
a_{11} & \ldots & a_{1n}\\
\dotfill & \dotfill & \dotfill\\
a_{mn} & \ldots & a_{mn}
\end{array}\right],\quad
\mathbf x = \left[\begin{array}{c} x_1\\  \vdots\\ x_n
\end{array}\right],\quad \mathbf b = \left[\begin{array}{c} b_1\\
\vdots \\ b_m\end{array}\right],\end{equation} we have the system
in the matrix form
\begin{equation}A\mathbf x = \mathbf b,\end{equation}
with the augmented matrix $\tilde{A} = [A|\mathbf b]$.

It is well known in linear algebra that by using elementary
transformations we can reduce the augmented matrix of the system
to the so called {\it Reduced Row Echelon Form (RREF)} and after
reenumerate the variables we can rewrite it in the form
\begin{equation} \RREF([A|\mathbf b]) = \left[\begin{array}{ccccccc}
1 &\ldots & 0 & \widetilde{a_{1,r+1}} & \ldots & \widetilde{a_{1,n}} & \widetilde{b_1}\\
\vdots & \ddots & \vdots & \ddots & \vdots & \vdots & \vdots\\
0 &\ldots & 1 & \widetilde{a_{r,r+1}} & \ldots & \widetilde{a_{r,n}} & \widetilde{b_r}\\
0 & \ldots & 0 & 0 & \ldots & 0 & 0\\
\dotfill &\dotfill & \dotfill &\dotfill &\dotfill &\dotfill &\dotfill
\end{array}\right].\end{equation}

The solutions of the system is some affine $m$-flat, $m=n-r$, $r=
\rk\widetilde{A}$ with a vector basis
\begin{equation}\left[\begin{array}{c} -\widetilde{a_{1,r+1}}\\ \vdots\\ -\widetilde{a_{r,r+1}}\\ 1\\ 0\\ \vdots\end{array}\right],\left[\begin{array}{c} -\widetilde{a_{1,r+2}}\\ \vdots\\ -\widetilde{a_{r,r+2}}\\ 0\\ 1\\ \vdots\end{array}\right],\ldots, \left[\begin{array}{c} -\widetilde{a_{1,n}}\\ \vdots\\ -\widetilde{a_{r,n}}\\ 0\\ \vdots\\ 1\end{array}\right]\end{equation}
and an affine point
\begin{equation}\left[\begin{array}{c}
\tilde{b_1}\\ \vdots\\ \tilde{b_r}\\ 0\\ \vdots\\
0\end{array}\right].\end{equation} In the particular case, where
$r=n=N$ the system has a unique solution. Start from here we
consider the only this nondegenerate case. The procedure of
producing the RREF of the augmented matrix
$\widetilde{A}=[A|\mathbf b]$ is written as some computer program.
We refer the readers to \cite{ptvf} for a program in C language.
It is easy to show that this computation needs at least the time
of order $2^N$. The most time complexity is paid to find a pivotal
element.

In this paper we try to simulate this procedure on quantum computers, and show that we could have the complexity time of order $2^{N/2}$.

The paper is organized as follows. We state our QGJE Algorithm and the Main Theorem in \S2. In \S3 we analyze the structure of quantum addition and explain its mechanism of work. Its seems to the authors that by using the representation theory and the corresponding Fast Fourier Transformation is easy to explain the quantum adder and also quantum multiplication. In \S5 we prepare the main background of quantum computing, and finally in \S5 the main theorem is proved.

\section{QGJE Algorithm}
\begin{enumerate}
\item[Step 1] Use the Grover's Search algorithm to find out  the
first non-zero $a_{i1}\ne 0$. \item[Step 2] If the search is
successful, produce the first leading $\mathbf 1$ in the first
place as $a_{11}$, else change to the next column and repeat step
1. \item[Step 3] Eliminate all other entries $a_{1,1},\dots,
a_{N,1}$ in the column. \item[Step 4] Change $N$ to $N-1$, control
if still $N>0$, repeat the procedure from the step 1. \item[Step
5] In backward eliminate all $a_{N-1,N}, \dots, a_{1,N}$.
\item[Step 6] Check if $N>0$, change $N$ to $N-1$ and repeat the
step 5.
\end{enumerate}
\begin{thm}
In the Quantum Gau\ss-Jordan Elimination Algorithm one needs at
most
\begin{equation}\frac{N(N-1)(2N+1)}{3} +
\left[\sqrt{2}\frac{(\sqrt{2})^N -1}{\sqrt{2}-1}\right] \sim
O(2^{N/2}) \end{equation} operations.
\end{thm}
Proof of this theorem is the main goal of the paper.

\section{A structural comment of Quantum Adder construction}
In this section we give a brief review of quantum computers, as we need in the rest of this paper.
\subsection{$q$-digit counting system}
Let us first remind the ordinary rule of arithmetic operations.
Let $p$ be a prime and $r$ some positive integer.
Write a number in series of $q=p^r$ with coefficients in the finite field $\mathbf F_q$,
\begin{equation}a = \sum_{i=0}^m a_iq^i, \qquad\qquad b=\sum_{i=0}^n b_jq^j, a_i,b_j\in \mathbf
F_q.\end{equation} The sum $a+b$  and the product $a.b$ are
defined as usually
\begin{equation}a+b = (a_0 + a_1q + \dots) + (b_0 +b_1q + \dots) = c_0 + c_1q + \dots,\end{equation}
\begin{equation}a.b= (a_0+a_1q+\dots)(b_0 + b_1q+\dots) = d_0+ d_1q+ ,\dots\end{equation}
where
\begin{equation}c_0 = a_0+b_0 (\mod  q),\end{equation}
\begin{equation}c_1 = a_1+b_1 +[(a_0+b_0)/q] (\mod\; q),\end{equation}
$$\dots\dots\dots\dots \dots$$
\begin{equation}d_0 = a_0.b_0 (mod\; q),\end{equation}
\begin{equation}d_1 = a_1.b_1 + [(a_0.b_0)/q] (\mod\; q),\end{equation}
$$\dots\dots\dots\dots\dots$$

\subsection{Qubits}
A fundamental notion of quantum computing is the notion of {\it
qubit}.
\begin{defn}{\rm
A {\it qubit} is a quantum system the Hilbert space of its quantum
states is the q-dimensional complex plane $\mathbf C^q$, and
therefore a quantum state of a qubit is a normalized complex
vector $\mathbf v\in \mathbf C^q$,
\begin{equation}\mathbf v = \left[\begin{array}{c} v_0\\ \vdots\\ v_{q-1}\end{array}\right], ||\mathbf v|| = |v_0|^2 +\ldots + |v_{q-1}|^2 = 1.\end{equation}
}\end{defn} Choose the standard basis of $q$ vectors
\begin{equation}|0\rangle =\left[\begin{array}{c}
1\\ 0\\ \vdots\\ 0\end{array}\right], |1\rangle =
\left[\begin{array}{c} 0 \\ 1\\ \vdots\\ 0 \end{array}\right],
\dots, |q-1\rangle = \left[\begin{array}{c} 0\\ 0\\ \vdots\\ 0\\
1\end{array}\right].\end{equation} An arbitrary vector can be
decomposed in this basis as a linear combination
\begin{equation}\mathbf v = v_0|0\rangle + v_1|1\rangle + \dots +v_{q-1}|q-1\rangle.\end{equation}
\begin{lem}
Every $n$-digit integer number can be written as a tensor product
of $n$ qubits \begin{equation}a = a_0 \otimes \dots \otimes
a_{n-1} \in (\mathbf C^2)^{\otimes n}, a_i \in \mathbf F_q=
\mathbf Z/q\mathbf Z.\end{equation}
\end{lem}
{\sc Proof.} Every integer can be written in binary form
\begin{equation}a= a_0 + a_12 + \dots + a_n2^n, a_i \in \mathbf F_q =\{ 0, 1,\dots,q-1 \}.\end{equation}
Therefore we have $|a_0\rangle \otimes \dots \otimes |a_n\rangle
\in (\mathbf C^2)^n.$ \hfill$\Box$

Let us consider the additive groups $G= (\mathbf F_q)^n$. A unitary character is a continuous homomorphism $\chi: G \to \mathbf S^1 = \mathbf U(\mathbf C)$ from the group $G$ into the group of unitary automorphism of $\mathbf C$, i.e. a continuous map $\chi: G \to \mathbf S^1 \cong \mathbf U(\mathbf C)$, such that
\begin{equation}\left\{\begin{array}{rcl}
\chi(x+y) &=& \chi(x)\chi(y)\\
\chi(0) &=& 1\\
\Vert \chi(x) \Vert &\equiv& 1.
\end{array}\right.\end{equation}
\begin{lem}
Every (multiplicative) unitary character of the additive groups $G=(\mathbf F_q)^n$ if of the form
\begin{equation}\chi_a(x) = \exp({\frac{2\pi i}{q^n}\sum_{i=1}^n a_ix_i}).\end{equation}
\end{lem}
{\sc Proof.} The lemma is an easy exercise from the group representation theory.\hfill$\Box$

It is convenient to write the number in fractional form, i.e. write
\begin{equation}0.a = \frac{a_0}{q^n} + \frac{a_1}{q^{n-1}} + \dots .\end{equation}
The integers modulo $2^n$ can be written as
\begin{equation}\bar{k} \mapsto e^{2\pi i 0.k}, \mbox{ where } 0.k = k/q^n,\end{equation}
for $k=0,1,\dots, q^n-1$. Let us consider the Haar measure $\mu(k)
= \frac{k}{q^n}$. Then we have the natural {\it Fast Fourier
Transform} (FFT).
\begin{defn}{\rm
The transformation \begin{equation}\FFT : |y\rangle \mapsto
\frac{1}{2^n}\sum_{k=0}^{q^n-1} e^{2\pi i
y_k\frac{k}{q^n}}|k\rangle\end{equation} is called the {\it Fast
Fourier Transform} of $|y\rangle$ }\end{defn} After that we can
formally write a number $|y\rangle$ in the form of $\exp(2\pi i
0.y_{n}y_{n-1}\dots y_0)$, i.e. we have
\begin{equation}\FFT : |y\rangle \mapsto
\exp(2\pi i 0.y_{n}y_{n-1}\dots y_0) \label{ffff} \end{equation}
\subsection{Quantum addition}
By using FFT we can write integers in the form \ref{ffff} i.e. we
have a 1-1 correspondence
\begin{equation}\CD a @>\FFT>> e^{2\pi i 0.a_{n}a_{n-1}\dots a_0}\endCD\end{equation}
\begin{equation}\CD b @>\FFT>> e^{2\pi i 0.b_{n}b_{n-1}\dots b_0}\endCD\end{equation}

\begin{defn}
The sum of two numbers $a$ and $b$ is defined as follows.
\begin{equation}\CD a @>\FFT>> e^{2\pi i0.a_n\dots a_0} @> \add\;  0.0\dots 0b_0 >>
e^{2\pi i0.a_na_{n-a}\dots a_1(a_0+b_0)} @>\add\; 0.0\dots 0b_1
>> \dots\endCD\end{equation}
\begin{equation}\CD \dots @> \add \; 0.b_n >> e^{2\pi i 0.(a_n+b_n)\dots
(a_0+b_0)} @>\FFT \inverse>> a+b\endCD \end{equation}
\end{defn}

\begin{rem} The reduction (memorize $[(a_i+b_i)/q]$ to
$a_{i+1} + b_{i+1}$ is the phase transition with the matrix
\begin{equation}\left[\begin{array}{cc}
1 & 0\\
0 & e^{2\pi i\frac{[(a_i+b_i)/q]}{q^{i-1}}}
\end{array}\right]\end{equation}
\end{rem}

\subsection{Quantum gates}
Let us recall some fundamental gates in quantum computing.

\subsubsection{Hadamard gate}
\begin{defn} The 1-qubit Hadamard gate $-\fbox{$\H$}-$ {\rm is defined by the
matrix}
\begin{equation}\H = \frac{1}{\sqrt{2}}\left[\begin{array}{cc}
1 & 1\\
1 & -1 \end{array}\right].\end{equation}
\end{defn}
{\rm The Hadamard gates acts on 1-qubits as follows}.
\begin{equation}\CD|x\rangle @>-\fbox{$\H$}->> \frac{1}{\sqrt{2}}\left((-1)^x|x\rangle + |1-x\rangle\right).\endCD\end{equation}

\subsubsection{Phase gate}
\begin{defn} The phase 2-qubit gate $-\fbox{$\Phi$}-$ is defined by the matrix
\begin{equation}\Phi = \frac{1}{\sqrt{2}}\left[\begin{array}{cc}
1 & 0\\
0 & e^{i\Phi}\end{array}\right].\end{equation} it produces an
action
\begin{equation}\CD |x\rangle @>-\fbox{$\Phi$}->> \frac{1}{\sqrt{2}}e^{i\Phi}|x\rangle.\endCD\end{equation}
\end{defn}

\subsubsection{CNOT(XOR) gate}
\begin{defn}{\rm
The 2-qubit CNOT (XOR)} gate {\rm is defined by the matrix}
\begin{equation}C = \left[\begin{array}{cccc}
1 & 0 & 0 & 0\\
0 & 1 & 0 & 0\\
0 & 0 & 0 & 1\\
0 & 0 & 1 & 0\end{array}\right].\end{equation} {\rm The action of
this gate is given by}
\begin{equation}\left\{\begin{array}{rcl}
|x\rangle & \to & |x\rangle\\
|y\rangle & \to & |x \oplus
y\rangle,\end{array}\right.\end{equation} {\rm where} $x\oplus y
:= x+ y \mod(2).$ \end{defn}

\subsubsection{Unitary gates}
\begin{defn}
The unitary gate -\fbox{$\U$}- {\rm is defined by a unitary matrix
U},
\begin{equation}\mathbf \U = \left[\begin{array}{cc}
1 & 0\\
0 & \U\end{array}\right].\end{equation}
\end{defn}

\subsubsection{Problem of quantum computing}
One of the major result concerning this problem is the following classification result.
\begin{thm}[Universal gates]
One-qubit gates, CNOT gate, phase gate and one of unitary gate generate all the other gates.
\end{thm}

\section{Quantum Grover's Search Algorithm}
\subsubsection{Black-box (Query) problem}
The problem is to compute some Boolean function of type
\begin{equation}f: \{0,1\}^n \to \{0,1\}.\end{equation}

\subsection{Deutsch's original problem}
The problem is to check whether the above defined Boolean function
is {\it balance}: the number of values 0's equals to the numbers
of values 1's, or {\it constant}.

\subsection{Search problem}
Given a Boolean function $f: \{0,1\}^n \to \{0,1\},$ defined by $f_k(x) = \delta_{xk}.$ Find $k$.

\subsection{Deutsch's scheme $-\fbox{H}-\fbox{f}-\fbox{H}-$}
The Deutsch's scheme produces the effects (modulo a constant, we
often omit it in order to not the formula longer, but it is easily
recovered in the case of necessary):
\begin{equation}\CD |x\rangle(|0\rangle -|1\rangle) @>-\fbox{H}-\fbox{f}-\fbox{H}->> \underbrace{(-1)^{f(x)}|x\rangle}_{\mbox{measurement}} (|0\rangle-|1\rangle).\endCD\end{equation}
In this scheme, the second qubit is of no interest, while the
first qubit has state \begin{equation}(-1)^{f(0)}|0\rangle +
(-1)^{f(1)}|1\rangle = \left\{\begin{array}{ll}
\pm(|0\rangle + |1\rangle), & \mbox{if } f=const\\
\pm(|0\rangle - |1\rangle), & \mbox{if } f=\mbox{\it
balanced}\end{array}\right.\end{equation}

\subsection{Quantum Grover's Search Algorithm}
Scheme: $-\fbox{$f_k$}-\fbox{H}-\fbox{$f_0$}-\fbox{H}-$.

This algorithm has special effect: Choose the state
\begin{equation}|S\rangle = \frac{1}{2^{N/2}}\sum_{i=1}^{N-1}|i\rangle,\end{equation}
and produce the Grover's iterate
\begin{equation}\CD |\psi\rangle(|0\rangle-|1\rangle) @>-\fbox{$f_k$}-\fbox{H}-\fbox{$f_0$}-\fbox{H}->> |\mbox{measurement}\rangle (|0\rangle -|1\rangle).\endCD\end{equation}

This iterate does nothing to any basis element $|i\rangle$ except
for $|k\rangle$ is changing into $-|k\rangle$, i.e. {\it the
reflection in the hyperplane perpendicular to the plane}
\begin{equation}\H|0\rangle =
|S\rangle =
\frac{1}{2^{N/2}}\sum_{i=1}^{N-1}|i\rangle,\end{equation} i.e.
rotation in the plane generate by $|k\rangle$ and $|S\rangle$ an
the Grover's iterate is a rotation of twice the angle from
$|k^\perp\rangle$ to $S^\perp\rangle$.

\section{Proof of the Main Theorem}
{\sc Step 1.} Let us consider the first column $\left[\begin{array}{c}
a_{11}\\ \vdots\\ a_{N1}\end{array}\right].$ Each $a_{i1}$ is either zero or non-zero. We do find the first non-zero value. In classical computation we needs to check all possible values of column entries and we needs $2^N$ operations. In quantum Grover's Search Algorithm we need to proceed $2^{N/2}$ operations. More precisely, choose a state
\begin{equation}|S\rangle = 2^{-N/2}\sum_{i=0}^{N-1} |i\rangle\end{equation} and produce the Grover's iterate
\begin{equation}\CD |\Psi\rangle(|0\rangle -|1\rangle) @>-\fbox{$f_k$}-\fbox{H}-\fbox{$f_0$}-\fbox{H}->> |\mbox{measurement}\rangle(|0\rangle- |1\rangle).\endCD\end{equation}
This iteration does nothing to any basis element except for
$|k\rangle$ is changing into $-|k\rangle$, i.e. reflection in the
hyperplane perpendicular to plane \begin{equation}\H|0\rangle =
2^{-N/2}\sum_{i=0}^{N-1} |i\rangle = |S\rangle,\end{equation} i.e.
a rotation in the plane generated by $|k\rangle$ and $|S\rangle$,
and the desired state $|k\rangle$ exactly is $\Phi$, the iterate
should be repeated $m$ times with
\begin{equation}(2m+1)\Phi \approx \frac{\pi}{2} \mbox{ or } m \approx \frac{\pi}{4\Phi} -\frac{1}{2}.\end{equation}
Since \begin{equation}\frac{1}{2^{N/2}} = \sin\Phi \approx \Phi,
\mbox{ we have } m \approx \frac{\Phi}{4}2^{N/2}.\end{equation}
This estimate is optimal in the sense that any another quantum
algorithm for searching an unstructured database must stake time
at least of order $O(2^{N/2})$.

Suppose we have found a non-zero element and denote it (the first nonzero element) by $a_{i1}$

{\sc Step 2.} Produce the first modified row
\begin{equation}\mathbf r_1 = [\mathbf 1,\frac{a_{i2}}{a_{i1}},
\dots, \frac{a_{iN}}{a_{i1}}].\end{equation} For this we need at
least $N-1$ multiplications with $\frac{1}{a_{i1}}$. Register the
resulted row as the first row.

{\sc Step 3.}
\begin{itemize}
\item Multiply the new first row with $-a_{j1}$. For this we need $N-1$ operations of multiplication and
\item Add the result to the $j^{th}$ row. For this we need another $N-1$ operations.
\end{itemize}
Totally in this step we need $2(N-1)^2$ arithmetic operations.

{\sc Step 4.} Now change $N$ to $N-1$ (need 1 subtraction) and control for the value $N>0$. We need one CNOT gate.

Loop back to the first step produce summation from 1 to $N$.

{\sc Steps 5 and 6.} Eliminating backward the numbers $a_{N-1,N},\dots, a_{1,N}$, we need $N-1$ additions and one CNOT gate to control and change $N$ into $N-1$.

{\sc All Steps.} Totally in all steps we need at most
\begin{equation}\sum_{n=1}^N (2^{n/2} + 2(n-1)^2 + 2(n-1) + 1 + 1)\end{equation}
operations.

{\sc Reduction.} The following lemma is trivial
\begin{lem}[Combinatorial Lemma]
\begin{equation}\sum_{n=1}^N n^2 = \frac{N(N+1)(2N+1)}{6}.\end{equation}
\end{lem}

{\sc Finish the Proof of the Main Theorem}
Denote the integral part of a real number $a$ by $[a]$.
Finally we have
\begin{equation}\sum_{n=1}^N(2^{n/2}+2(n-1)^2 +2(n-1) + 1 +1) \approx$$
$$2\sum_{n=1}^N n + 2\sum_{n=1}^N (n-1)^2 + [\sum_{n=1}^N \sqrt{2}^n] =$$
$$2\frac{N(N+1)}{2} + 2\frac{N-1)N(2N-1)}{6} + [\sqrt{2}\frac{\sqrt{2}^N-1}{\sqrt{2}-1}] =$$
$$\frac{(N-1)N(2N+1)}{3} + [\sqrt{2}\frac{\sqrt{2}^N-1}{\sqrt{23}-1}] \approx O(2^{N/2}).\end{equation}

\begin{cor}
For $N>>1$, the number of operations needed to produce the RREF of a square matrix using QGJE Algorithm is of order $O(2^{N/2})$.
\end{cor}
 In the next paper \cite{diepgiang} we apply this computation to  simulating of accounting principles on quantum computers.

\section*{Acknowledgments}
The work was partially reported in the Mathematical Club and Seminar of Topology and Geometry, Institute of Mathematics in Hanoi. We thank the people who attended for stimulating comments.

{\noindent\sc ${}^1$ Institute of Mathematics, Vietnam Academy of Science and Technology, 18 Hoang Quoc Viet Road, Cau Giay District, 10307 Hanoi, Vietnam\\
{\noindent\tt Email: dndiep@math.ac.vn}\\
\\
{\rm and}\\
\\
${}^2$ K47A1T, Department of Mathematics, University of Natural Sciences, Hanoi Vietnam National University, 40 Nguyen Trai, Thanh Xuan district, Hanoi, Vietnam}\\
{\noindent\tt Email: dhgiang84@gmail.com}
\end{document}